\documentclass[conference]{IEEEtran}
\usepackage{graphicx}
\usepackage{amsmath}
\usepackage{mathtools}
\usepackage{multirow}
\usepackage{booktabs}
\usepackage{rotating}
\usepackage{hhline}
\usepackage{textcomp}

\ifCLASSINFOpdf

\else

\fi

\begin{document}
	\title{Intra Prediction Using In-Loop Residual Coding for the post-HEVC Standard}

	\author{
		\IEEEauthorblockN{Mohsen Abdoli, F\'{e}lix Henry}
		\IEEEauthorblockA{Orange Labs\\
			Cesson S\'{e}vign\'{e}, France\\
			\{mohsen.abdoli, felix.henry\}@orange.com}
		\and
		\IEEEauthorblockN{Patrice Brault, Pierre Duhamel, Fr\'{e}d\'{e}ric Dufaux}
		\IEEEauthorblockA{L2S, CNRS - CentraleSupelec - Universit\'{e} Paris-Sud \\
			Gif-sur-Yvette, France\\
			\{patrice.brault, pierre.duhamel, frederic.dufaux\}@l2s.centralesupelec.fr}
	}
	
	\def\ps@IEEEtitlepagestyle{%
		\def\@oddfoot{\mycopyrightnotice}%
		\def\@evenfoot{}%
	}
	\def\mycopyrightnotice{%
		{\footnotesize\hfill The copyright belongs to me!\hfill}%   << here
		\gdef\mycopyrightnotice{}% just in case
	}
	
	\makeatletter
	%%%%%%%%%%%%%%%%%%%%%%%%%%%%%% User specified LaTeX commands.
	\def\ps@IEEEtitlepagestyle{%
		\def\@oddfoot{\mycopyrightnotice}%
		\def\@evenfoot{}%
	}
	\def\mycopyrightnotice{%
		%{\footnotesize\hfill 978-1-5090-3649-3/17/\$31.00 \textcopyright 2017 IEEE\hfill}%
		{\footnotesize\hfill 
			MMSP 2017 - IEEE 19$^{th}$ International Workshop on Multimedia Signal Processing - Luton, United Kingdom.
			\hfill}%
		\gdef\mycopyrightnotice{}% just in case
	}
	

	\maketitle
	
	\begin{abstract}
		A few years after standardization of the High Efficiency Video Coding (HEVC), now the Joint Video Exploration Team (JVET) group is exploring post-HEVC video compression technologies. In the intra prediction domain, this effort has resulted in an algorithm with 67 internal modes, new filters and tools which significantly improve HEVC. However, the improved algorithm still suffers from the long distance prediction inaccuracy problem. In this paper, we propose an In-Loop Residual coding Intra Prediction (ILR-IP) algorithm which utilizes inner-block reconstructed pixels as references to reduce the distance from predicted pixels. This is done by using the ILR signal for partially reconstructing each pixel, right after its prediction and before its block-level out-loop residual calculation. The ILR signal is decided in the rate-distortion sense, by a brute-force search on a QP-dependent finite codebook that is known to the decoder. Experiments show that the proposed ILR-IP algorithm improves the existing method in the Joint Exploration Model (JEM) up to 0.45\% in terms of bit rate saving, without complexity overhead at the decoder side.
	\end{abstract}
	
	\IEEEpeerreviewmaketitle
	\section{Introduction}
	\label{sec:intro}
	After two successful collaboration of ITU-T Video Coding Expert Group (VCEG) and ISO/IEC Motion Picture Expert Group (MPEG) in the standardization of the Advanced Video Coding (AVC) and the High Efficiency Video Coding (HEVC), a new collaboration team was formed in 2015 to investigate the potential need for the next generation video coding \cite{chen2015coding}. The agenda of this team that is called the Joint Video Exploration Team (JVET) is to study new video compression technologies for a future video coding standard. This effort has resulted in a state-of-the-art video codec called Joint Exploration Model (JEM). As of May 2017, JEM5.1 provides about 30\% bitrate saving compared to HEVC. Some of the most significant changes compared to HEVC are as follows:
	
	\begin{itemize}
		\item The Quad Tree Binary Tree (QTBT) block partitioning instead of the Quad Tree,
		\item Block size up to 256x256 instead of 64x64,
		\item 67 intra prediction modes (IPM) instead of 35,
		\item Multiple transforms,
		\item Adaptive motion vector resolution.
		
	\end{itemize}
	
	In the intra prediction, the new set of IPMs includes the traditional DC and planar plus 65 angular modes covering the same angle range as HEVC with double precision. Although the new intra prediction has a better performance than HEVC, it is still incapable to accurately predict when reference pixels from previous blocks are not well-correlated to the distant pixels inside the current block. This problem occurs when the content changes inside a block, which results in high energy concentration in the bottom-right corner of the residual block.
	
	The long distance prediction inaccuracy problem in lossy intra prediction has been sparsely studied in the literature. The Short Distance Intra Prediction (SDIP) \cite{cao2012short} was first proposed to Joint Collaborating Team on Video Coding (JCT-VC) in 2010 \cite{lai2010new}. In SDIP, the prediction remains at the Prediction Unit (PU) level by forcing a split into thin rectangular horizontal/vertical PUs, expecting that the small height/width of the PU solves the long distance prediction inaccuracy problem. In Combined Intra Prediction (CIP) \cite{dias2016improved}, the algorithm performs an additional inner-block pixel-level prediction and combines it with the traditional prediction using a weighted average. The fact that the pixel-level prediction of CIP uses an estimation of reconstructed inner-block references limits its performance, due to propagation of the estimation error. In \cite{gao2016differential}, the algorithm uses a DPCM-based pixel-level prediction with an out-loop residual signal. Like CIP, this method suffers from the estimation error propagation through the block. 
	
	In this paper, the proposed In-Loop Residual coding Intra Prediction (ILR-IP) tries to solve the long distance prediction inaccuracy problem by performing a pixel-level prediction using inner-block references. These inner-block references are reconstructed by the In-Loop Residual (ILR) signal. The ILR signal that has the same size as the block, is decided by a search loop in the rate-distortion sense and is transmitted besides the regular out-loop residual. Unlike the regular residual, the ILR transmission is done by vector quantization using a codebook that is known at the decoder side. As will be shown, using the ILR signal improves the prediction performance in case of in-block content change. 
	
	The rest of this paper is organized as follows. In section \ref{sec:principles}, the philosophy and general structure of ILR-IP is discussed. In section \ref{sec:algo}, the key elements of ILR-IP and different design choices are explained in detail. In section \ref{sec:result}, the performance of the proposed algorithm is discussed. Finally, section \ref{sec:conclusion} concludes the paper.
	
	\section{Principles of In-Loop Residual Intra Prediction (ILR-IP) }
	\label{sec:principles}
	
\begin{figure}
		\begin{center}
			\includegraphics[scale=0.4]{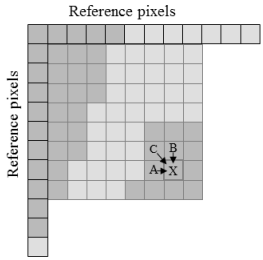}
			\caption{Long distance inaccuracy problem: regular references at top row and left column are not correlated to bottom-right area, due to the content change.}
			\label{fig:distantref}
		\end{center}
	\end{figure}

	The idea of inner-block pixel prediction for solving the long distance prediction inaccuracy problem can be better perceived at the decoder side, where reference pixels and residual signal are available. Figure \ref{fig:distantref} shows an example $8\times8$ block with its regular references, at the decoder side. According to the conventional intra prediction, a vertical angular IPM is most likely to be selected for this block, which copies reference pixels from the top reference line to all positions inside the block. As can be seen, the content in the bottom-right corner of the block is completely uncorrelated to its corresponding reference pixels. Normally, this would probably result in further splitting the block, thus requiring more data to be sent. However, it seems that the decoder might have the chance to use correlated inner-block local references in that corner without having to split the block. 
	
	For instance, at the time of pixel prediction at $X$, all surrounding pixels [$A,B,C$]  are already predicted by the regular intra prediction. Besides, the residual signal is available at the stage of decompression process, which makes [$A,B,C$] fully reconstructible. Therefore, it seems straightforward to utilize these inner-block local references for the prediction of $X$ at the decoder side. However the main obstacle is that, basically, the prediction decisions are made by an encoder which is unable to perform the same short distant prediction. The reason is that, unlike the decoder, the residual signal is not available at the stage of prediction at the encoder side. Therefore, it is impossible at the encoder side to reconstruct [$A,B,C$]. In other words, if the encoder aims at using [$A,B,C$] during the prediction step, it needs their residual values, which are calculated after the prediction step of the entire block is finished. Therefore there is a chicken-and-egg data dependency at the encoder side due to the block-based nature of the codec.  
	
	The ILR-IP algorithm removes the above chicken-and-egg data dependency, by providing the ILR signal, during the prediction at the encoder side. This signal is progressively added to predicted pixels to correct them and make them usable as inner-block references for the prediction of the following pixels within the current block. Since the main purpose of the ILR signal is the ``correction'' of predicted pixels, then the performance of the proposed algorithm directly depends on the pixel correction accuracy of the ILR signal. Therefore, the choice of the ILR signal is crucial at the encoder side. 
	
	In order to achieve the highest pixel correction accuracy, one might simply calculate a residual signal in the spatial domain that perfectly corrects pixels as if it were used as the ILR signal. This signal that is called the ``minimum residual (MinRes)'' in the remaining of this paper, can be easily calculated by accessing the original pixels at the encoder. In order to perform the same perfect pixel correction at the decoder, the MinRes signal needs to be transmitted in a lossless manner. However, this is not feasible due to its high rate. Therefore, a practical ILR signal must be decided in the rate-distortion sense. In the proposed algorithm, a vector quantization approach is used to train a codebook based on the MinRes values. This codebook is then fully explored in an encoder loop to select the ILR candidate with minimum rate-distortion cost (RDCost). The corresponding codebook index is then transmitted to the decoder.
	
	\section{Proposed ILR-IP algorithm}
	\label{sec:algo}
	
\begin{figure}
		\begin{center}
			\includegraphics[scale=0.4]{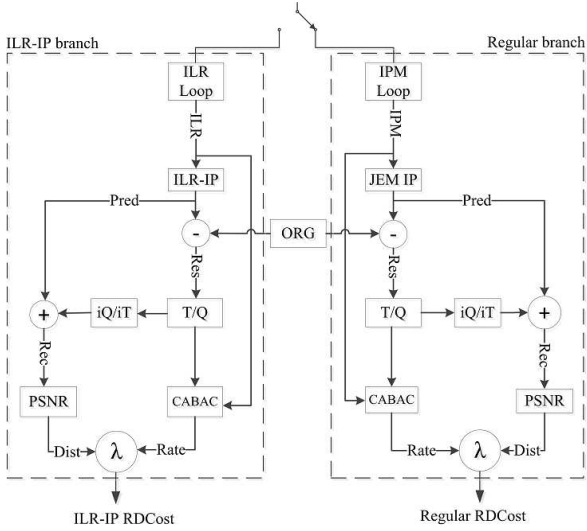}
			\caption{The block level decision based on the rate-distortion cost (RDCost) between the regular intra prediction and the ILR-IP.}
			\label{fig:diagram}
		\end{center}
	\end{figure}

	The existing intra prediction with 67 internal IPMs works very well for content with directional texture or plain structure, but has problems with more complex types of content. Therefore, the proposed ILR-IP is designed to coexist alongside the regular prediction, in order to improve the performance for problematic types of content. More specifically, in the new design, the intra prediction of each block has the choice between the regular intra prediction and the proposed ILR-IP algorithm. This choice is made in the rate-distortion sense and imposes an overhead of one flag per block to signal the selected algorithm. Figure \ref{fig:diagram} schematically shows the choice for the intra prediction algorithm at the block level. 
	
	In the ILR branch of Figure \ref{fig:diagram}, first the `ILR-Loop' box decides about the `ILR' signal. This decision can be made by any algorithm at the encoder side and needs to be transmitted to the decoder side. The ILR signal is then used as the input to the ILR-IP algorithm to produce the prediction signal (`Pred'). In the ILR-IP, the pixel prediction process starts from top-left corner of the block and progresses toward the bottom-right corner in a lexicographical scanning order. At each position of the scan, a short distance pixel prediction function takes three neighboring pixels (i.e. top, left, top-left) as input and predict the pixel at that position. The immediate step after the prediction is the pixel correction by adding the ILR value to the predicted value at that position. This step that will be expanded in the subsection \ref{subsec:2Dtestbed}, prepares the pixel that was just predicted, for future use as an inner-block reference. 
	
	Although the ILR signal is responsible for correcting the pixels after prediction, it is not yet able to completely replace the regular out-loop residual. Hence, the regular out-loop residual signal `Res' is also calculated by comparing the `Pred' signal to the original pixels values `ORG'. 
	
	As soon as all three signals of `ILR', `Pred' and `Res' are available, the RDCost of the block, after compression by ILR-IP, can be calculated. The rate and distortion calculations are performed in the `CABAC' and the `PSNR', respectively. For the rate calculation, the bits for transmission of the all syntax elements are counted. This includes the syntax elements of the ILR signal, the algorithm flag (to indicate use of the ILR-IP) and the regular out-loop residual. Since the ILR-IP does not utilize the regular IPMs, their corresponding syntax elements are obviously not encoded. 
	
	The current platform of ILR-IP is flexible in the sense that different algorithms can be implemented in the following component independently: 
	\begin{itemize}
		\item ILR signal decision: what values should be given to the algorithm as the ILR signal?
		\item ILR signal transmission: how should the ILR signal be transmitted to the decoder?
		\item Pixel prediction function: given the ILR signal, how pixels should be predicted using inner-block references?
	\end{itemize}
	
	%\section{Implementation and Experimental results}
	
	The existing intra prediction has been through a comprehensive optimization process which includes the adaptation of other codec tools that have an impact on/from the intra prediction. Hence, the performance of any new algorithm with major changes cannot be fairly evaluated, unless either all the tools are re-optimized, or the comparison is performed in an environment which is free from the impacts of other tools. In the rest of this section, first a fair simplified 1D experiment for validating the idea of ILR-IP is explained. Then a 2D implementation in the latest version of the JEM is provided.
	
	\subsection{1D experiment}
	\label{subsec:1Dtestbed}
	The main goal in this 1D experiment is to keep all the conditions as fair as possible for both regular and proposed intra prediction algorithms. The common specifications of this simplified 1D testbed include:
	\begin{itemize}
		\item The picture partitioning of one block size of 1$\times$4,
		\item The pixel prediction that takes two reference pixels and performs linear extrapolation (planar),
		\item Entropy coding of the regular quantized coefficients of the [out-loop] residual,
		\item The transform blocks by the 1D-DCT on 1$\times$4 blocks,
		\item One flag per block that is transmitted to signal the selected intra prediction algorithm.
	\end{itemize}
	
	\begin{figure}
		\begin{center}
			\includegraphics[scale=0.4]{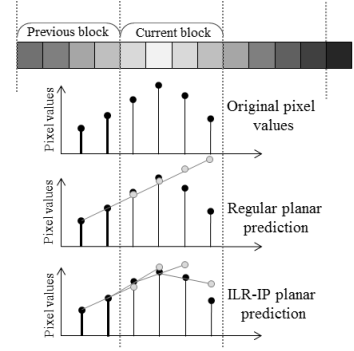}
			\caption{The regular and the ILR  intra prediction with planar pixel prediction function in the 1D testbed}
			\label{fig:1D}
		\end{center}
	\end{figure}
	
	Figure \ref{fig:1D} shows a line of image in the above 1D testbed, with a content change that coincides with the second 1$\times$4 block (i.e. current block). Here, the first two values in the diagram (i.e. the bold lines) are the regular reference pixels from the previous block. As can be seen, these references are not able to properly predict the last two pixels of the current block. However, the ILR-IP algorithm (which is given a proper ILR signal) is able to adapt to the change and predict the pixels by using corrected inner-block references. 
	
	According to the above specifications, the regular intra prediction algorithm is straightforward. However for the ILR-IP, two design choices have to be made:
	
	\subsubsection{ILR signal decision by brute force search in the transform domain} As explained, the MinRes in the spatial domain can be defined for each block, as a signal that provides zero distortion if used as the ILR signal. This signal can simply be calculated for each pixel, by 1) using original pixels as the inner-block prediction references, 2) predicting the pixel, and then 3) subtracting the predicted pixel from the original pixel.  
	
	Despite the fact that the MinRes signal is too costly to be transmitted in a lossless manner, the ILR-IP uses its quantized coefficients as the starting point for a brute-force search to obtain the 1D ILR signal. 
	Given the 1$\times$4 block of quantized coefficients of the MinRes signal in the transformed domain as $\textbf{T}_M=[t_M^{(1)},t_M^{(2)},t_M^{(3)},t_M^{(4)}]$, the ILR-IP finds the optimal ILR signal by exploring all candidates $\textbf{T}_C=[t_C^{(1)},t_C^{(2)},t_C^{(3)},t_C^{(4)}]$ where:
	\begin{equation}
	\label{eq:bruteforce1D}
	t^{(i)}_C=t^{(i)}_M+d^{(i)}, \text{  } i\text{=1,2,3,4}.
	\end{equation}
	
	The displacement $d^{(i)}$ of each coefficient in Eq. \ref{eq:bruteforce1D} is limited to integer values in an arbitrary interval $[-b,b]$. In the experiments presented in this section, $b=10$ is selected.
	
	\subsubsection{ILR signal transmission by entropy coding of the coefficient} Since the ILR signal is determined in the quantized transform domain, it is decided that the encoding method is the same as that of the regular out-loop residual. This includes using entropy coding for encoding the transform coefficients.
	
	Experiments show that the compression performance of the ILR-IP within the 1D testbed is better than the regular intra prediction by 6.2\% on average. These experiments are performed on the sequences of the classes B, C and D. One should note that the simplifications of the 1D testbed have highly impacted both performances. Hence, this compression gain is simply not achievable in the actual 2D JEM framework. However, it shows the potential of the proposed algorithm.
	
	\subsection{2D implementation in the state-of-the-art JEM video codec}
	\label{subsec:2Dtestbed}
	The ILR-IP algorithm in 2D is implemented inside the reference software JEM5.0.1. Here we introduce three main design choices made for the 2D ILR-IP model:
	\begin{figure}
		\begin{center}
			\includegraphics[scale=0.15]{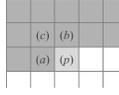}
			\caption{Three references (\textit{a}),(\textit{b}),(\textit{c}) around the pixel (\textit{p}) to be predicted.}
			\label{fig:predictionfun}
		\end{center}
	\end{figure}
	\subsubsection{Pixel prediction by LOCO-I function } The 2D pixel prediction function takes three neighbor pixels as references and implements the LOCO-I function \cite{weinberger2000loco}. This function, which is mostly used for lossless image compression (e.g. JPEG-LS), checks the existence of a vertical or horizontal edge. Eq. \ref{eq:prediction} formulates the LOCO-I in Figure. \ref{fig:predictionfun} at prediction position (\textit{p}), given the three references at positions (\textit{a}),(\textit{b}),(\textit{c}). 
	\begin{multline}
	\label{eq:prediction}
	\hat{x}^{(p)}=f(\tilde{x}^{(a)}, \tilde{x}^{(b)}, \tilde{x}^{(c)})=
	\\
	\begin{cases}
	min(\tilde{x}^{(a)},\tilde{x}^{(b)}), & \text{if } \tilde{x}^{(c)} \leq max(\tilde{x}^{(a)},\tilde{x}^{(b)}).\\
	max(\tilde{x}^{(a)},\tilde{x}^{(b)}), & \text{if } \tilde{x}^{(c)} \geq min(\tilde{x}^{(a)},\tilde{x}^{(b)}).\\
	\tilde{x}^{(a)}+\tilde{x}^{(b)}-\tilde{x}^{(c)}, & \text{Otherwise.}
	\end{cases}
	\end{multline}
	
	Generally, $\tilde{x}^{(s)}$ at scanning position (\textit{s}) can either be a ``corrected'' pixel inside current block or a ``decoded'' pixel from previous blocks. In the case that $\tilde{x}^{(s)}$ belongs to current block (i.e. inner-block reference), the correction is done by: 
	
	\begin{equation}
	\label{eq:correction}
	\tilde{x}^{(s)}=\hat{x}^{(s)}+y^{(s)},
	\end{equation}
	where $y^{(s)}$ is the ILR value at (\textit{s}) and will be discussed later.
	\subsubsection{ILR signal decision by vector quantization}
	\label{subsec:vectorquant}
	Unlike the 1D experiment, the brute-force search approach is not practical in 2D, due to the very large search space. Therefore, the ILR signal in 2D is determined by a vector quantization approach with a finite QP-dependent codebook \cite{huang2015mode}. 
	
	Ideally, the codebook should provide proper ILR signals for all block, close enough to their MinRes. Therefore, we decide to adopt the MinRes idea to train the codebook by a modified version of the iterative Linde-Buzo-Gray (LBG) algorithm \cite{linde1980algorithm}. To do so, a training sequence consisting of many H$\times$W video blocks is prepared. These blocks contain actual content to be compressed by ILR-IP and are taken from a large and diverse set of video sequences with different resolutions and contexts (i.e. natural, synthetic etc.). In addition, an initial codebook is randomly set. This choice is not critical as the goal is to gradually converge to a proper codebook by iterating with the LBG.  Each iteration of the LBG-based codebook construction consists of two steps, namely ``classification'' and ``update'':
	
	\textit{Classification}: Given a codebook at iteration $t$, represented by $N$ centroids $C^t=\{\textbf{Y}_i^t;i=1,2,...,N\}$, as well as a large set of training sequences $\textbf{S}$, the classification step decides about the centroid of each sample $\textbf{X} \in \textbf{S}$ and categorizes it in one of the $N$ classes:
	\begin{equation}
	\label{eq:classification}
	R^t_i=\{\textbf{X}:d(\textbf{X},\textbf{Y}_i^t) < d(\textbf{X},\textbf{Y}_j^t); \text{all } j \in [1,N] \text{ and } j \neq i\},
	\end{equation}
	where $d(\textbf{X},\textbf{Y})$ is an arbitrary function for calculating the distance between the sample \textbf{X} and the centroid candidate \textbf{Y}. Here, categorizing block \textbf{X} into class \textit{i} means selecting the centroid $\textbf{Y}_i$ as the ILR signal of \textbf{X}. 
	
	The standard LBG is normally used for representing a large set of points by a limited set of centroids and the goal is to minimize the approximation error from the centroids. Hence, a $\rho$-norm distance measure (i.e. Euclidean distance) is used as the distance function $d$. However, in the context of video compression, we aim at minimizing the RDCost as \textit{d}, rather than a $\rho$-norm measure. 
	
	\textit{Update}: After classification of all samples, each centroid of the codebook is updated by a centroid function that basically minimizes a global distance measure based on \textit{d}:
	\begin{equation}
	\label{eq:update}
	C^{t+1}=\{\mathrm{cent}(R^t_i);i=1,...,\mathrm{N}\}.
	\end{equation}
	
	In the standard LBG, the arithmetic mean as the centroid function, along with the Euclidean distance as \textit{d}, guarantee a minimization of the global distance measure. Moreover, enough iterations on Eq. \ref{eq:classification} and Eq. \ref{eq:update} guarantee the convergence of the codebook \cite{el2016design}. However, the codebook construction problem in the ILR-IP is slightly more complicated, as all decisions are made through a non-linear prediction process. More precisely, the spatially progressive prediction scheme of ILR-IP makes it theoretically possible to decrease prediction error, while increasing Euclidean distance from the MinRes. As a consequence, the arithmetic mean as the centroid function in Eq. \ref{eq:update} no longer guarantees the global distance minimization. 
	
	Due to the above limitation, a slight, yet important modification is made on the centroid function of the LBG algorithm. In the proposed method, each centroid is updated pixel-by-pixel. This is comparable to the block-level updating if the standard LBG was adopted. The difference is that in the pixel-level updating, when one value of a centroid is updated, it will immediately be used for updating the next values of that centroid. However, in the block-level updating, all values at all scanning positions are updated at once by using the old centroid from the previous iteration. In the rest of this section, all discussions for the centroid updating are only focused on the $i$-th class of the codebook ($i=1,2,...,N$).
	
	Let $R^t_i=\{\textbf{X}_l;l=1,...,L\}$ be the set of samples that were classified in the \textit{i}-th class at the \textit{t}-th iteration by Eq. \ref{eq:classification}. Each sample in this set is an H$\times$W block. Moreover, a scanning order specifies the order of predicting pixels as well as updating centroid values. Therefore, given the scanning order and the training samples of the $i$-th class as $R^t_i$, the \textit{l}-th sample can be vectorized following the scanning order as:
	
	\begin{equation}
	\textbf{X}_l=[x^{(1)}_l,...,x^{(\textrm{H} \times \textrm{W})}_l].
	\end{equation}
	
	The corresponding centroid for the above samples at iteration \textit{t}, can accordingly be vectorized in the same order:
	
	\begin{equation}
	\textbf{Y}_i^t=[y_i^{t,(1)},...,y_i^{t,(\textrm{H} \times \textrm{W})}].
	\end{equation}
	
	Moreover, the prediction pixels may be calculated by applying Eq. \ref{eq:prediction} given $\textbf{Y}_i^t$:
	
	\begin{equation}
	\hat{\textbf{X}}_l^t=[\hat{x}^{t,(1)}_l,...,\hat{x}^{t,(\textrm{H} \times \textrm{W})}_l].
	\end{equation}
	
	Note that the superscript \textit{t} on each predicted pixel $\hat{x}$ indicates that its corresponding inner-block references at its top, left and top-left, were corrected by the centroid at iteration \textit{t}. By referring to Figure \ref{fig:predictionfun}, Eq. \ref{eq:prediction} and Eq. \ref{eq:correction}, this can be written:
	\begin{multline}
	\label{eq:correctpredict}
	\hat{x}^{t,(p)}_l=f(\tilde{x}^{t,(a)}_l,\tilde{x}^{t,(b)}_l,\tilde{x}^{t,(c)}_l)=\\
	f(\hat{x}^{t,(a)}_l+y^{t,(a)}_i,\hat{x}^{t,(b)}_l+y^{t,(b)}_i,\hat{x}^{t,(c)}_l+y^{t,(c)}_i).
	\end{multline}
	
	The adopted centroid function performs a pixel-level update with respect to the scanning order: 1) it starts from the first scanning position of the centroid at iteration $t$; 2) it updates the centroid value at that position; 3) it replaces the updated value in the ongoing centroid of iteration $t$. After these steps are done for all positions of the scanning order, the entire centroid is updated and ready for the next iteration $t+1$. Eq. \ref{eq:updateOurs} formulates the pixel-level update at position (\textit{p}) that requires performing the pixel prediction for all \textit{L} samples of the class at that position:
	\begin{equation}
	\label{eq:updateOurs}
	y^{t+1,(p)}_i=\dfrac{1}{L}\sum_{l=1}^{L}{(o^{(p)}-\hat{x}_l^{t+1,(p)})},
	\end{equation}
	where $o^{(p)}$ is the original pixel value at the scanning position (\textit{p}). As shown, $\hat{x}^{t+1,(p)}_l$ belongs to the iteration $t+1$ instead of $t$. According to Eq. \ref{eq:correctpredict}, this indicates that its inner-block references were corrected by the centroid of iteration $t+1$ that was just updated in the ongoing update phase. This is possible since the scanning positions at top, left and top-left of the (\textit{p}) are precedent to (\textit{p}), thus, their corresponding centroid values are already updated in the current update phase.
	
	After Eq. \ref{eq:updateOurs} is performed on all scanning positions (\textit{p}), the $i$-th centroid for the iteration $t+1$ is given as:
	\begin{equation}
	\textbf{Y}_i^{t+1}=[y_i^{t+1,(1)},...,y_i^{t+1,(\textrm{H} \times \textrm{W})}].
	\end{equation}
	
	The above steps explain the centroid update step in one iteration of the codebook construction process for one centroid of the codebook. The goal is to repeat both the classification and update steps until convergence. 
	
\begin{figure}
		\begin{center}
			\includegraphics[scale=0.55]{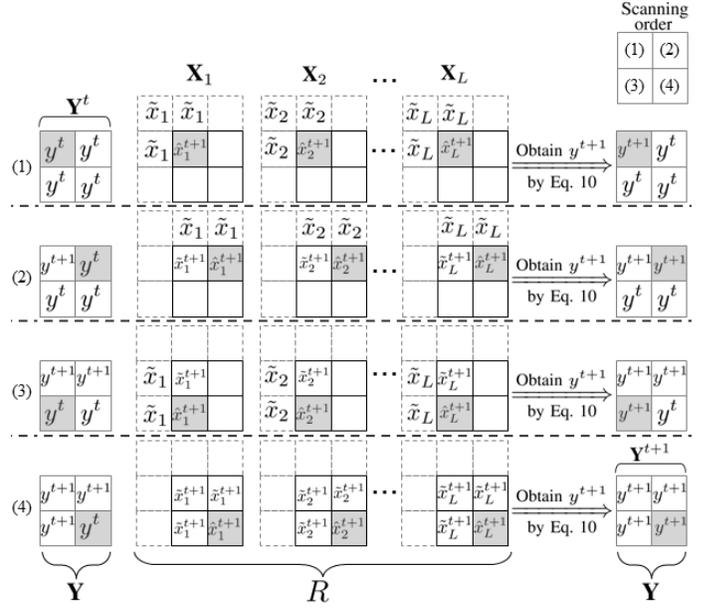}
			\caption{Four stages of pixel-level centroid update for a 2$\times$2 block according to the given scanning order. At each stage, the gray pixel indicates the scanning position of the centroid that is being updated. The regular references of the 2$\times$2 blocks are shown by dotted squares.}
			\label{fig:vqdiagram}
		\end{center}	
	\end{figure}

	Figure \ref{fig:vqdiagram} visually summarizes the above steps for a 2$\times$2 centroid \textbf{Y}, with \textit{L} samples in $R$ and the given scanning order. As can be seen, the update process starts with the $\textbf{Y}^{t}$ in the first row and first column. At the end of each row, the corresponding centroid value at that position is updated and replaced in the $\textbf{Y}^t$. This process repeats until the last position is updated and the whole centroid is updated to $\textbf{Y}^{t+1}$.
	\subsubsection{ILR signal transmission by fixed-length coding} Unlike the 1D testbed where entropy coding was used, here the fixed-length coding of codebook index is used  for IRL transmission. This is due to the probability distribution function of the centroids indexes that roughly follows uniform probability. Therefore, the ILR transmission simply includes equiprobable encoding of $n=\textrm{log}(N)$ bits indicating one of $2^n$ centroids in the codebook of size \textit{N}.

	\section{Compression performance of the 2D ILR-IP}
	\label{sec:result}
	%\label{subsec:results2D}
	In this section, the performance of the ILR-IP algorithm is compared to the state-of-the-art reference software of JEM5.0.1 in All-Intra coding mode and conforming the Common Test Conditions (CTC). Without loss of generality, here the algorithm is limited to 4$\times$4 blocks. Experiments are run with codebooks of size 16, 32, 64, 128 and 256. These codebooks are trained by the modified LBG algorithm introduced in section \ref{subsec:vectorquant} and on an independent training set.
	
	To reach the maximum performance of JEM, all the new tools except non-separable secondary transforms (NSST) \cite{zhao2016nsst} and explicit multiple core transforms (EMT) \cite{zhao2016enhanced} are activated. In fact these two tools are perfectly optimized for encoding the regular out-loop residual signal, while the shape of the out-loop residual of the ILR-IP blocks is essentially different due to the existence of the ILR. Consequently, the existing transform choices do not necessarily improve out-loop residual coding of the ILR-IP as efficiently as they improve the regular out-loop residual coding. Hence, these two transform-related tools are switched off in both algorithms for a fair comparison. 
	
	Table \ref{tab:bitsaving} shows the Bj$\o$ntegaard delta (BD) rate of the ILR-IP against JEM5.0.1, where it consistently outperforms the JEM. The peak performance of the ILR-IP is for the artificial videos in the Class F. This can be justified by the fact that in artificial videos, in-block content changes usually appear on sharp edges that separate two pure plain areas. Therefore, changes can be compensated more easily by ILR signals that have non-zero values only on the edges. Table \ref{tab:complexity} shows the run-time complexity comparison between the ILR-IP and the JEM5.0.1. As shown, the proposed ILR-IP algorithm brings compression gain with no complexity overhead at the decoder side, which is crucial for adoption of a new technology. 
	
	\begin{table}[htbp]
		\centering
		\caption{BD-rate of the ILR-IP compared to the JEM5.0.1 with five different codebook sizes (all values are in percentage \%)}
		\begin{tabular}{|c|p{1.9cm}|c|c|c|c|c|}
			\hhline{~~-----}
			\multicolumn{2}{c|}{}       & \multicolumn{5}{|c|}{Codebook size} \\ \hhline{~~-----}
			\multicolumn{2}{c|}{}       & 16    & 32    & 64    & 128   & 256 \\ \hhline{-------}
			\multirow{4}[0]{*}{\begin{sideways}Class A\end{sideways}} & Traffic & -0,34 & -0,32 & -0,32 & -0,32 & -0,33 \\ \hhline{~------}
			& PeopleOnStreet & -0,28 & -0,32 & -0,33 & -0,32 & -0,31 \\ \hhline{~------}
			& NebutaFestival & 0,02 & -0,01 & -0,01 & -0,02 & 0,01 \\ \hhline{~------}
			& SteamLocomotive & 0,02 & 0,00 & -0,01 & -0,01 & 0,01 \\ \hhline{-------}
			\multirow{5}[0]{*}{\begin{sideways}Class B\end{sideways}} & Kimono & -0,06 & -0,01 & -0,01 & 0,00 & -0,02 \\ \hhline{~------}
			& ParkScene & -0,22 & -0,25 & -0,25 & -0,21 & -0,26 \\ \hhline{~------}
			& Cactus & -0,14 & -0,15 & -0,18 & -0,19 & -0,20 \\ \hhline{~------}
			& BasketballDrive & 0,00 & -0,01 & -0,01 & 0,04 & 0,02 \\ \hhline{~------}
			& BQTerrace & -0,10 & -0,22 & -0,27 & -0,36 & -0,40 \\ \hhline{-------}
			\multirow{4}[0]{*}{\begin{sideways}Class C\end{sideways}} & BasketballDrill & 0,00 & -0,09 & -0,17 & -0,19 & -0,17 \\ \hhline{~------}
			& BQMall & -0,20 & -0,27 & -0,32 & -0,45 & -0,48 \\ \hhline{~------}
			& PartyScene & -0,16 & -0,22 & -0,30 & -0,39 & -0,43 \\ \hhline{~------}
			& RaceHorses & 0,02 & -0,06 & -0,09 & -0,08 & -0,09 \\ \hhline{-------}
			\multirow{4}[0]{*}{\begin{sideways}Class D\end{sideways}} & BasketballPass & -0,04 & -0,12 & -0,22 & -0,22 & -0,49 \\ \hhline{~------}
			& BQSquare & -0,16 & -0,26 & -0,45 & -0,63 & -0,78 \\ \hhline{~------}
			& BlowingBubbles & -0,16 & -0,20 & -0,35 & -0,24 & -0,25 \\ \hhline{~------}
			& RaceHorses & 0,03 & -0,13 & -0,06 & -0,27 & -0,25 \\ \hhline{-------}
			\multirow{3}[0]{*}{\begin{sideways}ClassE\end{sideways}} & FourPeople & -0,37 & -0,38 & -0,41 & -0,36 & -0,36 \\ \hhline{~------}
			& Johnny & -0,29 & -0,43 & -0,55 & -0,50 & -0,55 \\ \hhline{~------}
			& KristenAndSara & -0,24 & -0,47 & -0,68 & -0,76 & -0,70 \\ \hhline{-------}
			\multirow{4}[0]{*}{\begin{sideways}ClassF\end{sideways}} & BasketballDrillTx & -0,12 & -0,15 & -0,40 & -0,50 & -0,41 \\ \hhline{~------}
			& ChinaSpeed & -0,73 & -0,82 & -0,65 & -1,13 & -1,29 \\ \hhline{~------}
			& SlideEditing & -0,35 & -0,45 & -0,57 & -0,86 & -1,47 \\ \hhline{~------}
			& SlideShow & -0,74 & -0,83 & -1,21 & -1,51 & -1,66 \\ \hhline{-------}
			\multicolumn{1}{c|}{} & \textbf{Average} & \textbf{-0.20} & \textbf{-0.26} & \textbf{-0.32} & \textbf{-0.4} & \textbf{-0.45} \\ \hhline{~------}
		\end{tabular}%
		\label{tab:bitsaving}
	\end{table}%
	
	From Table \ref{tab:bitsaving}, it can be seen that the performance of the ILR-IP linearly improves as the size of the codebook increases. Therefore, one might aim at further increasing the codebook size to achieve better performance. However, the main obstacle in the current design is the encoder complexity of codebook exploration. Currently, the encoder has to calculate RDCost for all centroids in the codebook to pick the best one.  
	
	It is important to note that the current codebook exploration can be accelerated without major impact on the performance. For instance, one might explore only a small subset of the codebook to pick the suboptimal ILR signal for each block. It was observed during implementations that the second-best and third-best ILR candidates in the codebooks are not much worse than the optimal ILR. This is left as future work.

	\begin{table}[htbp]
		\centering
		\caption{Run-time complexity of the ILR-IP algorithm compared to the JEM5.0.1 at the encoder and the decoder side}
		\begin{tabular}{|l|c|c|c|c|c|}
			\hhline{~-----}
			\multicolumn{1}{c|}{} & \multicolumn{5}{c|}{Codebook size} \\
			\hhline{~-----}
			\multicolumn{1}{c|}{} & 16    & 32    & 64    & 128   & 256 \\ \hhline{------}
			ET (\%) & 117   & 126   & 137   & 165   & 216 \\ \hhline{------}
			DT (\%) & 100   & 100   & 100   & 100   & 100 \\ \hhline{------}
		\end{tabular}%
		\label{tab:complexity}
	\end{table}%

	\section{Conclusion}
	\label{sec:conclusion}
	In this paper, the long distance prediction inaccuracy problem was targeted. The main philosophy was that in case of in-block content change, the references from previous blocks are uncorrelated and therefore, nearer in-block pixels should be used as reference. The proposed ILR-IP introduces an additional signal to progressively correct the inner-block pixels during the prediction. The corrected pixels are then used as reference instead of the references from previous blocks. 
	
	The validity of the ILR-IP algorithm was first tested in a 1D testbed where it outperformed the regular prediction. In the 2D, the proposed algorithm was implemented in the software JEM5.01. The encoder was given the possibility pick the best algorithm according to the RDCost. This imposed an overhead of one extra flag per block plus the encoder side complexity. Experiments show that the ILR-IP is able to compensate the flag overhead and bring up to 0.45\% BD-rate saving, without complexity overhead at the decoder side.
	
	The current version of the ILR-IP is limited to the 4$\times$4 blocks and codebook size up to 256. In future works, we will consider extending it to larger blocks and codebooks. This will give ILR-IP more opportunity to improve the RDCost for high QPs where 4$\times$4 blocks are less likely to be selected.

	\bibliographystyle{unsrt}
	\bibliography{myBib}

\end{document}